\documentstyle[12pt,world_sci]{article}
\begin{document}
 
\title{{\bf NEW RESULTS FROM CLEO-II \\
            ON HADRONIC DECAYS OF THE TAU LEPTON\footnote{\rm
                Contributed talk presented at the Eighth Meeting
                of the Division of Particles and Fields, Albuquerque,
                New Mexico, August 2--6, 1994.}
      }}
\author{Jon Urheim \\
  {\em %Lauritsen Laboratory of High Energy Physics\\
     California Institute of Technology \\
     Pasadena, California USA 91125 \\ }
%\vspace{0.3cm}
\vspace*{0.3cm}
(Representing the CLEO collaboration)\\
{\em }}
 
\maketitle
\setlength{\baselineskip}{2.6ex}
 
\begin{center}
\parbox{13.0cm}
{\begin{center} ABSTRACT \end{center}
{\small \hspace*{0.3cm} 
Results on semi-hadronic decays of the $\tau$ lepton are presented, from 
studies of $e^+e^-$ annihilation data obtained at the Cornell Electron
Storage Ring with the CLEO-II detector.  
Branching fractions have been measured for decays to two, three
and four hadrons, namely $\tau^-\!\!\rightarrow\! \nu_\tau h^-\pi^0$,
$\tau^-\!\!\rightarrow\! \nu_\tau h^-h^+h^-$, and 
$\tau^-\!\!\rightarrow\! \nu_\tau h^-h^+h^-\pi^0$, where $h^\pm$ represents a 
charged pion or kaon.  CLEO-II has also observed decays with charged and/or 
neutral kaons; preliminary results for branching ratios and structure
arising from the decay dynamics are given.  Connections
are made with predictions derived from theoretical models, 
the Conserved Vector Current theorem, isospin constraints and sum rules.
}}
\end{center}
 
\section{Introduction}
 
Semi-hadronic decays of the $\tau$ lepton provide a unique opportunity to 
explore the hadronic charged weak current.  
The leptonic current is well-understood in electroweak theory, and 
the $\tau$ is the only lepton which can decay semi-hadronically. 
Thus, one can probe the weak and strong dynamics governing the 
formation of hadronic final states.  

\begin{figure}
  \vspace{5.cm}
  \includegraphics{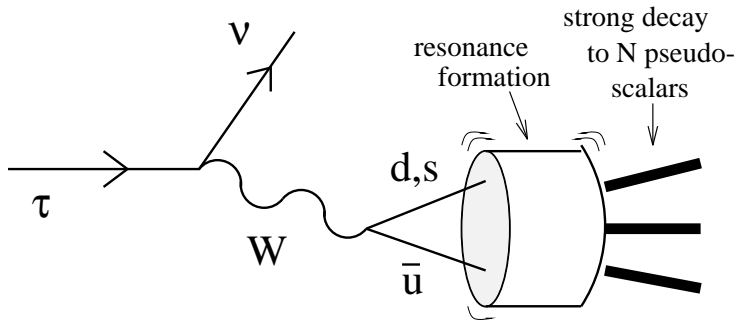}
  \caption[]{\small Simplified picture of semi-hadronic $\tau$
             lepton decay.}
\label{fig:cartoon}
\end{figure}

A cartoon of semi-hadronic $\tau$
decay is shown in Figure \ref{fig:cartoon}.
The values of $q^2$ carried by the $W$ are small, so the strong
physics is non-perturbative, dominated by resonance formation, and gives
rise to final states consisting of light pseudoscalar mesons.  
%which may only decay weakly or electromagnetically.

Spin, parity, isospin and charge conjugation symmetries constrain the
hadronic states which can form \cite{Tsai}.  
The spin and parity quantum numbers of
a given state select out either the vector or axial vector part of the
weak V--A current.  Isospin and charge conjugation for Cabibbo-favored 
($\bar{u} d$) currents require that vector states produced be G-parity 
even and axial vector states be G-parity odd, and further restrict
them to decay to even and odd numbers of pions respectively.
While there is little other theoretical input regarding $\tau$ decays
to axial vector states, production of non-strange vector states can be 
related to low energy $e^+e^-$ cross-sections by way of the 
Conserved Vector Current (CVC) theorem.  Also sum-rules can be invoked
to relate the vector and axial-vector (or strange and non-strange) 
hadronic spectral functions (of $q^2$) which describe the resonant 
content of the hadronic system.

The present experimental situation is somewhat murky however.  
%Nearly twenty years since their first observation, 
Only a few $\tau$ decays have been studied with
sufficient precision to test the elements of the hadronic weak and
strong physics described above.  Inconsistencies are present, both
in regard to the overall picture of $\tau$ decay and within
the data for several specific decays.
Below, I present results from CLEO-II on semi-hadronic
$\tau$ decay branching fractions.  Comments on resonant structure
are made where appropriate.  
%Unless otherwise indicated, all results are preliminary.  
This talk is organized as follows:
A brief description of the CLEO-II detector is given in
Section \ref{s-detector}.  In Section \ref{s-inc}, results are given
for decays containing from 2 to 5 hadrons.  
%--- without separating charged pions and kaons.  
Results on decays containing one or two kaons are
discussed in Section \ref{s-kaons}.

\section{The CLEO-II Detector}
\label{s-detector}

CLEO-II \cite{cleo} is a solenoidal spectrometer and calorimeter operating at 
the south interaction region of the Cornell Electron Storage Ring (CESR),
where $\tau$ leptons are produced in pairs through $e^+e^-$ annihilation
at $\sqrt{s}\sim 10.6$ GeV, with a $\sim 0.91$ nb cross section.  Since
the turn-on of CLEO-II in late 1989, CESR has delivered in excess of 4 
fb$^{-1}$ of integrated luminosity, corresponding to $\sim 3.6$ million 
$\tau$-pairs.  The components of CLEO-II most relevant to the analyses 
presented here are the 67-layer tracking system and the 7800-element
CsI(Tl) crystal calorimeter which are contained within a 1.5T magnetic
field, and cover $>96\%$ of $4\pi$ in solid angle.  The segmentation and
energy resolution of the calorimeter permit reconstruction of 
$\pi^0\!\!\rightarrow\! \gamma\gamma$ decays with high efficiency for $\pi^0$'s
of all energies: typical $\pi^0$ mass resolutions range from 5 to 7 MeV/c$^2$,
and we have observed $\tau$ decays containing as many as four fully 
reconstructed $\pi^0$'s \cite{npi0}.  Muons above $\sim\! 1$ GeV/c are 
tracked in Iarocci tubes deployed at three depths in the flux return steel.
Limited $\pi/K$ separation is achieved below 1 GeV/c and above 2 GeV/c
using $dE/dx$ information from the 51-layer main drift chamber and/or
time of flight measurements from scintillation counters located in 
front of the calorimeter.

%%%%%%%%%%%%%%%%%%%%%%%%%%%%%%%%%%%%%%%%%%%%%%%%%%%%%%%%%%%%%%%%%%%%%%%%%%%
%%%%%%%%%%%%%%%%%%%%%%%%%%%%%%%%%%%%%%%%%%%%%%%%%%%%%%%%%%%%%%%%%%%%%%%%%%%
 
\section{Decays to Two, Three, Four and Five Hadrons}
\label{s-inc}

CLEO-II has directed much effort to precisely measure branching 
fractions for decays to charged hadrons with and without
accompanying $\pi^0$'s.  This focus was motivated by the persistence
of the one-prong deficit \cite{gilman,pdg}. Earlier work includes studies 
of $\tau$ decays to one charged hadron plus two or more $\pi^0$'s
\cite{npi0} as well as decays containing $\eta$ mesons \cite{eta}.
We observed for the first time the decay 
$\tau^- \!\!\rightarrow\! \nu_\tau 2\pi^- \pi^+ 2\pi^0$ \cite{piompi0}, 
finding a surprisingly large branching fraction ($0.48\pm 0.04 \pm 0.04 \%$).
Below I present results for the decays 
$\tau^-\!\!\rightarrow\! \nu_\tau h^- \pi^0$, 
$\tau^-\!\!\rightarrow\! \nu_\tau 2h^-
h^+ [\pi^0]$ and $\tau^-\!\!\rightarrow\! \nu_\tau 3h^- 2h^+ [\pi^0]$.
In these analyses, $h^\pm$ represents a charged hadron ($\pi$ or $K$)
which we do not attempt to identify.

%%%%%%%%%%%%%%%%%%%%%%%%%%%%%%%%%%%%%%%%%%%%%%%%%%%%%%%%%%%%%%%%%%%%%%%%%%%
\subsection{The Decay $\tau^-\!\!\rightarrow\! \nu_\tau h^- \pi^0$}

The primary semi-hadronic $\tau$ decay,
$\tau^-\!\!\rightarrow\! \nu_\tau \pi^-\pi^0$, occurs through the 
vector portion of the hadronic charged weak current.  The
$\rho(770)$ meson dominates the spectral function responsible for this
decay, but contributions from excited $\rho$'s and
non-resonant $\pi^-\pi^0$ formation may also be present.  

The CVC theorem allows one to relate the $\tau^-\!\!\rightarrow\!
\nu_\tau\pi^-\pi^0$ decay width to the iso-vector contribution to the 
$e^+e^-\!\!\rightarrow\! \pi^+\pi^-$ cross section.  A branching fraction
of $24.58\pm 0.93\pm 0.27\pm 0.50\%$ is predicted \cite{KS,marc},
where the errors are from uncertainties
in the $e^+e^-$ data, $\tau$ lifetime, and knowledge of the radiative
corrections.  Experimental values range from 22\% to 26\%
\cite{pdg}, limited in precision by difficulties in 
$\pi^0$-finding and rejection of decays with multiple neutrals 
(themselves poorly measured until recently).

We have studied the decay $\tau^-\!\!\rightarrow\! \nu_\tau h^- \pi^0$ 
\cite{brho}, $h^- \equiv \pi^-, K^-$, using a 1.58 fb$^{-1}$ 
data sample ($1.44\times 10^6$ $\tau$-pairs).  We employ three 
statistically independent methods, determining
the branching fraction, $B_{h\pi^0}$, according the following
forms:
\begin{equation}
  B_{h\pi^0} = \sqrt{\frac{N_{e-h\pi^0} N_{\mu-h\pi^0}}
                    {2 N_{e-\mu}
                    N_{\tau\tau} }} , \quad
                \sqrt{\frac{N_{h\pi^0-h\pi^0}}{N_{\tau\tau}}}, 
                 \quad \hbox{or} \quad
                \frac{N_{3h-h\pi^0}}{N_{3h-1}} \;  B_1.
\label{eq-brho}
\end{equation}
The $N$'s denote the measured background-subtracted, efficiency-corrected 
event yields for the given final states ($e$ denotes 
$\nu e\bar{\nu}$, {\em etc.}).
%$3h$ denotes $\nu hhh[\pi^0]$, $1$ denotes inclusive decay to one charged
%particle, and we denote $\nu h\pi^0$ by $\rho$ for notational brevity).  
$N_{\tau\tau}$ is the number of $\tau$-pairs, as determined from the 
integrated luminosity (known to 1\% from studies of Bhabha scattering,
$e^+e^-\!\!\rightarrow\! \gamma\gamma$, and 
$e^+e^-\!\!\rightarrow\!\mu^+\mu^-$),
and the theoretical cross-section (also known to 1\%).
$B_1$ is the world average topological one-prong branching fraction.

Due to space constraints, 
we merely give the results below, and refer the reader 
elsewhere \cite{brho} for experimental details.  
The three methods agree; combining them we find:
\begin{equation}
 B_{h\pi^0} = 25.87 \pm 0.12 \,({\rm stat.}) \pm 0.42 \,({\rm syst.})\, \%,
\end{equation}
based on samples totalling $\sim 44,000$ 
reconstructed $\tau^-\!\!\rightarrow\!\nu_\tau h^- \pi^0$ decays.  
The dominant systematic errors (common to all three methods)
are those due to $\pi^0$-finding efficiency (0.9\%) and the veto on 
events with additional energy deposition in the calorimeter applied to 
reject backgrounds (0.9\%).  

This result is higher than most other measurements 
\cite{pdg}, but is consistent with recent values obtained at LEP 
by OPAL \cite{opalrho}, L3 \cite{l3rho}, and Aleph \cite{alephrho} which
also tend to be high.  Subtracting the measured contribution from
$\tau^-\!\!\rightarrow\! \nu_\tau K^-\pi^0$,\cite{ys} we find 
$B_{\pi^\pm \pi^0} 
=25.36\pm 0.44\%$, in good agreement with the CVC prediction.

%%%%%%%%%%%%%%%%%%%%%%%%%%%%%%%%%%%%%%%%%%%%%%%%%%%%%%%%%%%%%%%%%%%%%%%%%%%
\subsection{The Decays $\tau^-\!\!\rightarrow\! \nu_\tau h^- h^+ h^- [\pi^0]$}

As with the decay described above, our understanding of $\tau$ decay into 
final states with three charged particles has been confused by
conflicting experimental data \cite{pdg}.  
The CLEO-II observation of decays containing two $\pi^0$'s with a 
branching fraction around 0.5\% \cite{piompi0} has implications for
3-prong results from other experiments.  Below, I present 
preliminary results from CLEO-II on branching fractions for the decays 
$\tau^-\!\!\rightarrow\! \nu_\tau h^-h^+h^-$ and
$\tau^-\!\!\rightarrow\! \nu_\tau h^-h^+h^-\pi^0$.

The decay $\tau^-\!\!\rightarrow\! \nu_\tau h^-h^+h^-$ is expected to occur
through the axial-vector current.  Hence theoretical constraints are
limited.  The hadronic system is expected to be dominated by the 
$a_1(1260)$, which gives rise to both $\pi^-\pi^+\pi^-$ and 
$\pi^-\pi^0\pi^0$ final states.  However channels with charged kaons are
also known to contribute.

Because the decay $\tau^-\!\!\rightarrow\! \nu_\tau h^-h^+h^-\pi^0$ has an even
number of final state mesons, the vector current is expected to play the 
primary role.  Due to severe Cabibbo and phase-space suppression for 
modes containing kaons, the four-pion final state is expected to 
dominate.  Thus, predictions can be made from $e^+e^-\!\!\rightarrow\! 4\pi$
data through application of the CVC theorem.  Two such calculations 
give the branching fraction to be $4.3\pm 0.3\%$ \cite{EI} and 
$4.8\pm 0.7\%$ \cite{NP}.  Experimental results cluster around the 5\%
level.  

At CLEO-II, branching fractions are obtained from analyses of the following
event types: $3h^\pm$--$3h^\pm$, $\ell$--$3h^\pm$ and $\ell$--$3h^\pm\pi^0$,
where $\ell$ represents an electron or muon \cite{glas3h}.  
The branching fractions
are determined from measured quantities as follows:
\begin{equation}
  B_{3h^\pm}  =  \sqrt{\frac{N_{3h^\pm-3h^\pm}}{N_{\tau\tau}}} 
                 \quad {\rm or} \quad
                 \frac{N_{\ell-3h^\pm}}{N_{\tau\tau}B_\ell}, 
  \quad \quad
  B_{3h^\pm\pi^0} = \frac{N_{\ell-3h^\pm\pi^0}}{N_{\tau\tau}B_\ell}, 
\label{eq-b3h}
\end{equation}
where as before the $N$'s represent background-subtracted, 
efficiency-corrected event yields, $N_{\tau\tau}$ is the number of
produced tau-pairs ($1.44\times 10^6$), and $B_\ell$ represents the
leptonic branching fraction.  
%Due of the large branching fraction for 
%$\tau^-\!\!\rightarrow\! \nu_\tau h^-h^+h^-$, 
%there is sufficient statistics
%in the $3h^\pm$--$3h^\pm$ sample that despite the much smaller number 
The much smaller $3h^\pm$--$3h^\pm$ sample 
(2,030 events as opposed to 17,479 in the $\ell$--$3h^\pm$ sample) 
gives a nearly comparable statistical error because of the square 
root.  Also systematic errors associated with event-related
efficiencies are smaller by a factor of 2 than in the 
$\ell$--$3h^\pm$ analysis. 

\begin{figure}
% \vspace{6.cm}
  \begin{picture}(450,175)(0,0)
    \put(380,135){(c)}
  \includegraphics{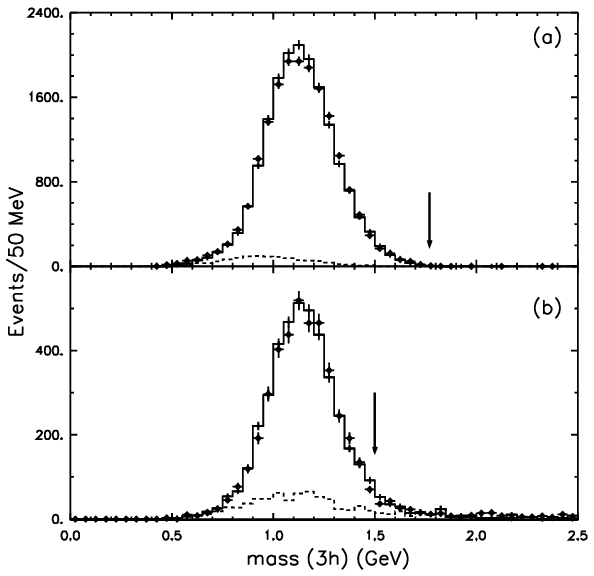}
  \includegraphics{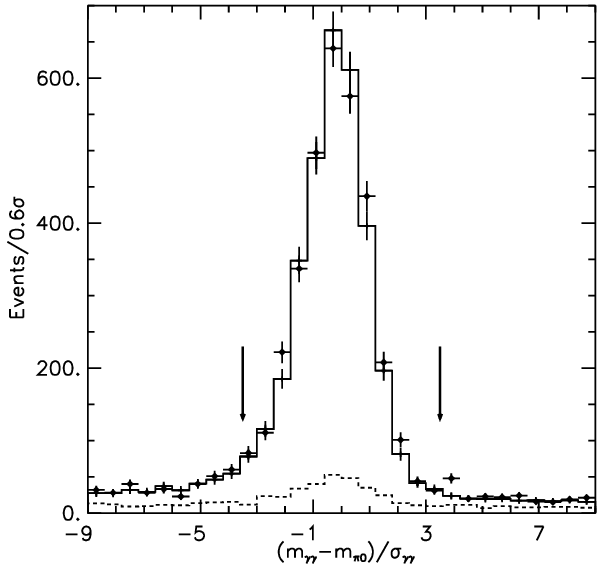}
  \end{picture}
  \caption[]{\small Invariant mass spectra for $\tau^-\!\!\rightarrow\!
             \nu_\tau h^-h^+h^-$ decays, taking $M_h = M_\pi$, 
             for lepton-tagged events (a) and $3h^\pm$--$3h^\pm$ (b)
             event samples.  In (c) is shown the quantity 
             $(M_{\gamma\gamma}-M_{\pi^0})/\sigma_{M_{\gamma\gamma}}$
             for events in the lepton tagged $\tau^-\!\!\rightarrow\! \nu_\tau
             h^-h^+h^-\pi^0$ candidate sample.
             The data (points) are shown along with
             the expectations for signal plus background from the Monte
             Carlo \cite{koralb,geant} (solid histogram).  
             Also shown are the expected
             contributions from backgrounds (dashed histogram).
             Arrows indicate the location of cuts.
            }
\label{fig:m3h}
\end{figure}

In Figure \ref{fig:m3h}(a,b), we plot the invariant mass spectra for the 
hadronic system in the $\tau\!\!\rightarrow\! \nu_\tau h^-h^+h^-$ 
event samples, taking the pion mass for the hadrons.  
These samples are relatively background-free as demonstrated by the
scarcity of events with $M_{3h} > M_\tau$:  non-$\tau$ backgrounds are
estimated to comprise $4.7\pm 0.9 \%$ in the $3h^\pm$--$3h^\pm$
sample and only $0.2\pm 0.2\%$ in the $\ell$--$3h^\pm$ sample.  In 
Figure \ref{fig:m3h}(c), the presence of $\pi^0$'s is demonstrated 
in events containing two energy clusters unassociated with the charged
tracks.

The combined results obtained from these event 
samples are:
\begin{eqnarray}
 B_{3h^\pm} 
      & = & 9.82\pm 0.09\,({\rm stat.})\pm 0.34\,({\rm syst.})\, \% \\
 B_{3h^\pm\pi^0} 
      & = & 4.25\pm 0.09\,({\rm stat.})\pm 0.26\,({\rm syst.})\, \%.
\end{eqnarray}
While the statistical errors are significantly smaller than those previously
attained, the systematic errors are comparable to or slightly worse than
recent reports from LEP \cite{tomasz}.  Uncertainties in modelling 
track-finding efficiency (2--2.5\%), in effects of cuts on
unaccounted-for energy deposition applied to suppress backgrounds (3--4\%),
and in the $\pi^0$-reconstruction efficiency for $3h^\pm\pi^0$ events (3\%) 
dominate the systematic errors.  These errors are preliminary, and 
they should shrink after further study.

The $\tau^-\!\!\rightarrow\! \nu_\tau h^- h^+ h^-$ result is significantly 
higher than the PDG '94 average value of $8.0\pm 0.4\%$ \cite{pdg}, supporting 
the recent preliminary value reported by Aleph ($9.57\pm 0.24\pm 0.22 
\%$).  After correcting for modes containing kaons, CLEO-II results 
in both channels \cite{npi0,brho} indicate 
$B_{3\pi^\pm}\approx B_{\pi^\pm 2\pi^0}$, as expected.

Correspondingly, the $\tau^-\!\!\rightarrow\! \nu_\tau h^- h^+ h^- \pi^0$
result is somewhat lower than previous results, although still agreeing
with the CVC predictions noted earlier.  Summing the two 
results along with the result for $B_{3h^\pm 2\pi^0}$ 
\cite{piompi0} and accounting for common errors, we obtain 
a total 3-prong $\tau$ branching fraction of $14.55\pm 0.13\pm 0.59\%$.
This is consistent with the PDG average value of $14.32\pm 0.27\%$ 
\cite{pdg} for the topological branching fraction.

%%%%%%%%%%%%%%%%%%%%%%%%%%%%%%%%%%%%%%%%%%%%%%%%%%%%%%%%%%%%%%%%%%%%%%%%%%%
\subsection{Five-Prong Tau Decays}

CLEO-II has measured branching fractions for decays into final
states with five charged particles \cite{fivepi}.  The results are
\begin{eqnarray}
   B(\tau^-\!\!\rightarrow\!\nu_\tau 3\pi^-2\pi^+ 
            + \geq 0\,{\rm neutrals}) & = & 0.097\pm 0.005\pm 0.011\,\% \\
   B(\tau^-\!\!\rightarrow\!\nu_\tau 3\pi^-2\pi^+) 
                                      & = & 0.077\pm 0.005\pm 0.009\,\% \\
   B(\tau^-\!\!\rightarrow\!\nu_\tau 3\pi^-2\pi^+\pi^0) 
                                      & = & 0.019\pm 0.004\pm 0.004\,\%, 
\end{eqnarray}
the most precise to date.

%%%%%%%%%%%%%%%%%%%%%%%%%%%%%%%%%%%%%%%%%%%%%%%%%%%%%%%%%%%%%%%%%%%%%%%%%%%
%%%%%%%%%%%%%%%%%%%%%%%%%%%%%%%%%%%%%%%%%%%%%%%%%%%%%%%%%%%%%%%%%%%%%%%%%%%

\section{Decays to Final States Containing Kaons}
\label{s-kaons}

The production of final states containing kaons represents
a largely unexplored sector of $\tau$ lepton decay.  In the following
sections, I present results \cite{ys,glask}
from CLEO-II on decays with one, two, or
three final state mesons of which at least one is a kaon.

%%%%%%%%%%%%%%%%%%%%%%%%%%%%%%%%%%%%%%%%%%%%%%%%%%%%%%%%%%%%%%%%%%%%%%%%%%%
\subsection{The Decay $\tau^-\!\!\rightarrow\! \nu_\tau K^-$}

The decay $\tau^-\!\!\rightarrow\!\nu_\tau K^-$ is completely specified in
electroweak theory through analogy with kaon decay.  The branching ratio, 
accounting for radiative effects, is predicted to be $0.74\pm 0.01\%$ 
\cite{marc}.  CLEO-II has measured it to be
$0.66\pm 0.07\pm 0.09 \%$ \cite{ys}.

%%%%%%%%%%%%%%%%%%%%%%%%%%%%%%%%%%%%%%%%%%%%%%%%%%%%%%%%%%%%%%%%%%%%%%%%%%%
\subsection{Decays of the Type $\tau^-\!\!\rightarrow\! \nu_\tau [Kh]^-$}

Like the decay $\tau^-\!\!\rightarrow\! \nu_\tau\pi^-\pi^0$, decays of the
type $\tau^-\!\!\rightarrow\! \nu_\tau [Kh]^-$ are expected to occur through
the vector part of the weak current.  The $K^- K^0$ final state has zero net 
strangeness, and CVC predictions based on $e^+e^-\!\!\rightarrow\! \pi^+\pi^-$
data and SU(3) symmetry give branching fractions of $0.11\pm 0.03\%$
\cite{EI} and $0.16\pm 0.02$ \cite{NP}.  A recent analysis by Aleph 
\cite{alephk0} gives $B_{K^-K^0} = 0.29\pm 0.12\pm 0.03\%$.
The $K^-\pi^0$ and $\pi^-\bar{K}^0$ final states are expected to appear in
the ratio $1:2$ from isospin, with a spectral function dominated
by the $K^*(892)$ resonance.  In this case, the Das-Mathur-Okubo (DMO) sum 
rule \cite{DMO} allows one to predict $B_{[\pi K]^-}$ from the 
measured value of $B_{\pi^-\pi^0}$ discussed earlier.

At CLEO-II, all three final states ($K^-K^0$, $K^-\pi^0$ and $\pi^-\bar{K}^0$)
have been observed, using $dE/dX$ (and time-of-flight for $K^-\pi^0$) for 
charged kaon identification, and detection of $K^0_S \!\!\rightarrow\! 
\pi^+\pi^-$ decay for neutral kaons.  The $K^-\pi^0$ decay branching fraction
has been measured with limited statistics to be $0.51\pm 0.10\pm 0.07\%$, 
in the analysis \cite{ys} described in the previous section.
A higher-statistics study of the $\pi^-\bar{K}^0$ decay based on
2.96 fb$^{-1}$ of data ($2.71\times 10^6$ $\tau$-pairs) is described
below, in which the $K^-K^0$ decay is also extracted.

\begin{figure}
% \vspace{6.cm}
  \begin{picture}(450,175)(0,0)
    \put(50,135){(a)}
    \put(210,135){(b)}
    \put(340,135){(c)}
  \includegraphics{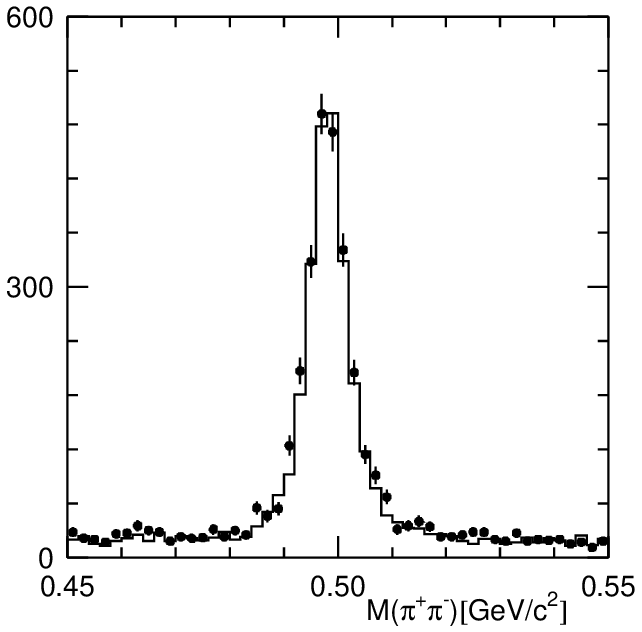}
  \includegraphics{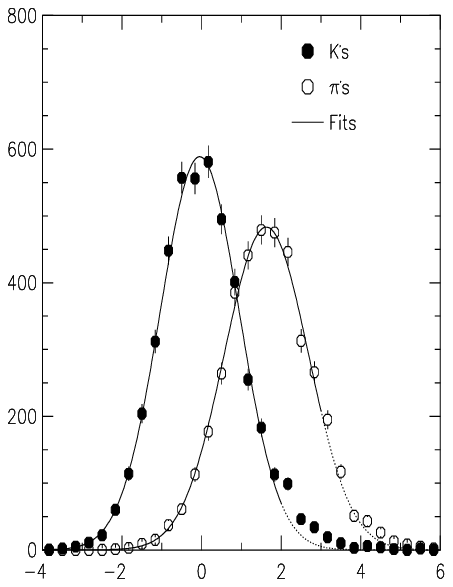}
  \includegraphics{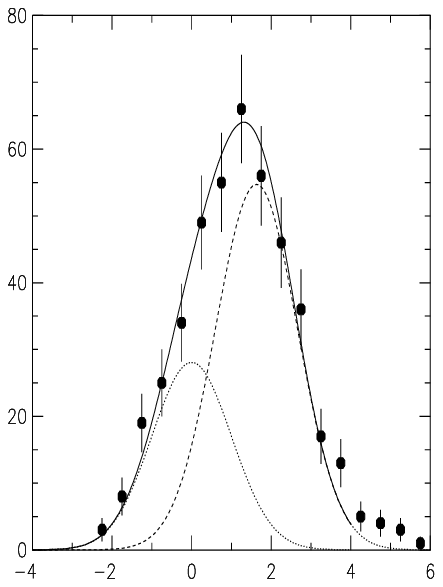}
  \end{picture}
  \caption[]{\small (a) Invariant mass of $\pi^+\pi^-$ pairs forming
             detached vertices in the
             $\tau^- \!\!\rightarrow\! \nu_\tau h^- K^0_S$ candidate data 
             (points) and Monte Carlo (histogram) samples.  (b) Plots
             of $\sigma_K$ (see text) for $\pi$'s and $K$'s with momenta
             above 2 GeV/c, identified from kinematics in ${D^*}^+
             \!\!\rightarrow\! D^0 \pi^+\!\!\rightarrow\! 
             (K^-\pi^+)\pi^+ $ decays.
             (c) Plot of $\sigma_K$ for the charged hadron accompanying
             the $K^0_S$ in $\tau^-\!\!\rightarrow\! \nu_\tau h^- K^0_S$, where
             $p_{h}>2$ GeV/c, fit to a sum (solid curve) of Gaussians
             parameterizing the $dE/dx$ response to pions (dashed) and
             kaons (dotted) taken from the $D^*$ data shown in (b). 
            }
\label{fig:ksh}
\end{figure}

We select events of a 1--3 charged-track topology in which two oppositely
charged tracks of the `3' form a vertex separated from the $e^+e^-$ 
interaction point by at least 5 mm.
In Figure \ref{fig:ksh}(a), the presence of $K^0_S\!\!\rightarrow\! 
\pi^+\pi^-$ decays is demonstrated by plotting the invariant mass of the 
detached tracks as pions.  
After background subtraction, we have $1,938\pm 126$ candidates for 
the decay $\tau^-\!\!\rightarrow\! \nu_\tau h^- K^0_S$.  

For momenta above 2 GeV/c, $dE/dx$ gives $\sim 2\sigma$ $\pi/K$ separation, 
as shown in \ref{fig:ksh}(b) for kinematically selected $\pi$'s and $K$'s
from $D^0\!\!\rightarrow\! K^-\pi^+$ data.  
Selecting $\tau^-\!\!\rightarrow\! \nu_\tau h^- K^0_S$
events with $p_h> 2$ GeV/c, we plot in Figure \ref{fig:ksh}(c)
the difference between the measured $dE/dx$ 
and that expected for kaons, divided by experimental resolution,
(denoted $\sigma_K$), and fit to a linear combination of response
functions for $\pi$'s and $K$'s.

\begin{figure}
% \vspace{6.cm}
  \begin{picture}(450,175)(0,0)
    \put(50,145){(a)}
  \includegraphics{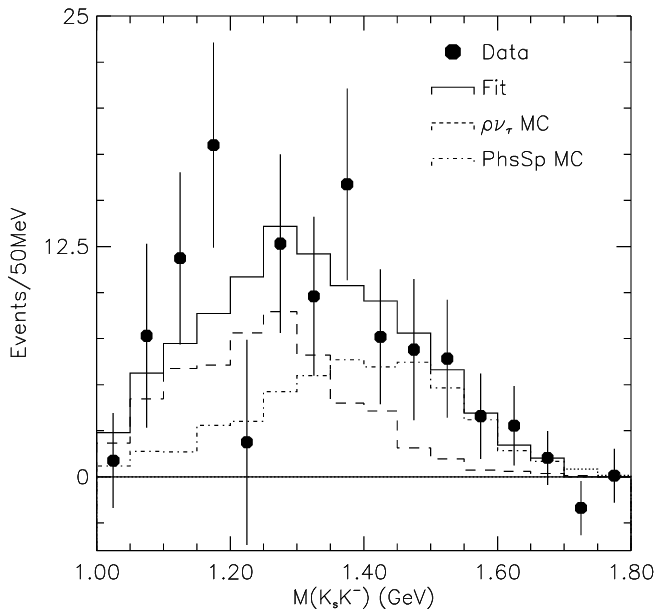}
  \includegraphics{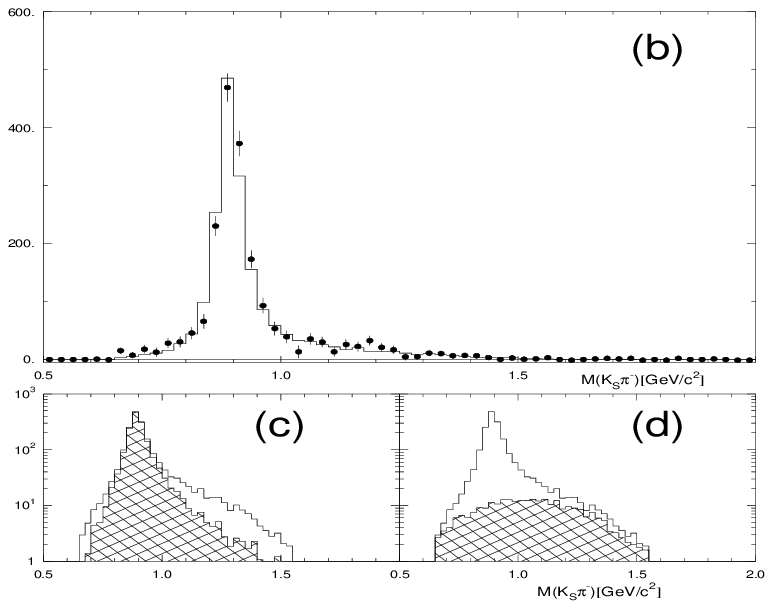}
  \end{picture}
  \caption[]{\small (a) Invariant mass of $K^- K^0_S$ system for
             $\tau^-\!\!\rightarrow\! \nu_\tau K^- K^0_S$ candidate events
             in the data (points), along with a fit (solid histogram)
             to $\rho^-$ (dashed) and phase space (dotted) contributions
             as determined through Monte Carlo simulation. (b) Plot
             of $M_{\pi^- K^0}$ for  $\tau^-\!\!\rightarrow\!
             \nu_\tau h^- K^0_S$ candidates in the data (points),
             along with a sum (solid histogram) of distributions obtained
             from $\tau^-\!\!\rightarrow\! \nu_\tau {K^*}^-(892)$ and 
             $\tau^-\!\!\rightarrow\! \nu_\tau K^- K^0_S$ 
             Monte Carlo samples (shown separately in (c) and (d), 
             respectively), where the latter is determined according to
             the mixture of $\rho$ and phase-space constributions 
             shown in (a).
            }
\label{fig:mksh}
\end{figure}

We obtain the following preliminary branching fractions from these studies:
\begin{eqnarray}
   B(\tau^-\!\!\rightarrow\!\nu_\tau h^-K^0) 
                                      & = & 0.977\pm 0.023\pm 0.108\,\% \\
   B(\tau^-\!\!\rightarrow\!\nu_\tau K^-K^0) 
                                      & = & 0.123\pm 0.023\pm 0.023\,\%, 
\end{eqnarray}
The result for $B_{K^-K^0}$ is lower than that from Aleph, and
consistent with expectations from CVC.  In Figure \ref{fig:mksh}, 
we plot the invariant masses of the $K^-K^0_S$ and $h^-K^0_S$ systems
in these event samples.  
The $K^-K^0_S$ system is softer than that predicted
by a phase-space model, and may have some resonant content such as in the
`$\rho$' model, based on the $\tau^-\!\!\rightarrow\!\nu_\tau\pi^-\pi^0$ 
spectral function.  The $h^-K^0_S$ system, 
after accounting for the admixture of $K^-K^0_S$ events, 
is consistent with saturation by 
$\tau^-\!\!\rightarrow\!\nu_\tau {K^*}^-(892)$,
as shown in Fig.~\ref{fig:mksh}(b), permitting a test of the DMO sum
rule.  Using the CLEO-II results on $h^-\pi^0$, $K^-\pi^0$, $h^-K^0$ and
$K^-K^0$ final states we find $B_{K^*}/B_{\rho} = (0.98\pm 0.03\pm 0.13)
\times \tan^2{\theta_C}$, where the DMO sum rule prediction is 
$0.93 \, \tan^2{\theta_C}$ in the narrow-width approximation.

%%%%%%%%%%%%%%%%%%%%%%%%%%%%%%%%%%%%%%%%%%%%%%%%%%%%%%%%%%%%%%%%%%%%%%%%%%%
\subsection{Decays of the Type $\tau^-\!\!\rightarrow\! \nu_\tau [Kh\pi]^-$}

\begin{figure}
% \vspace{6.cm}
  \begin{picture}(450,175)(0,0)
    \put(390,55){(g)}
  \includegraphics{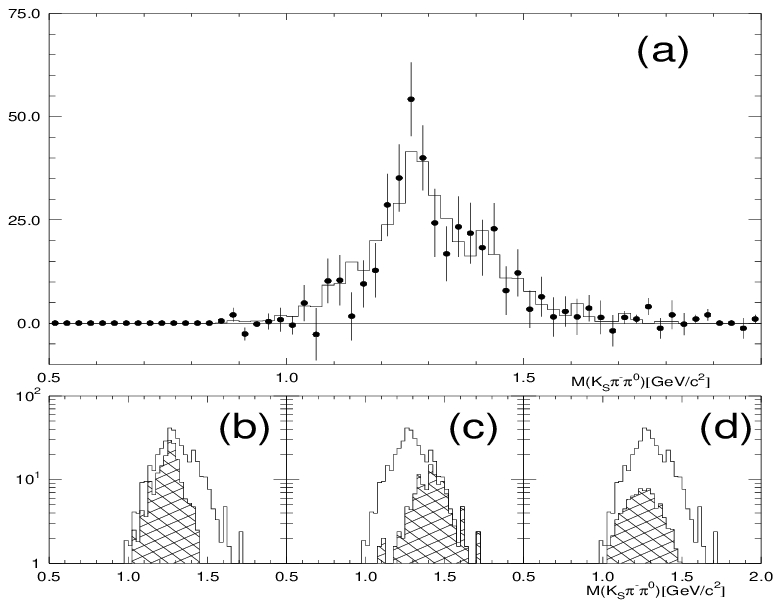}
  \includegraphics{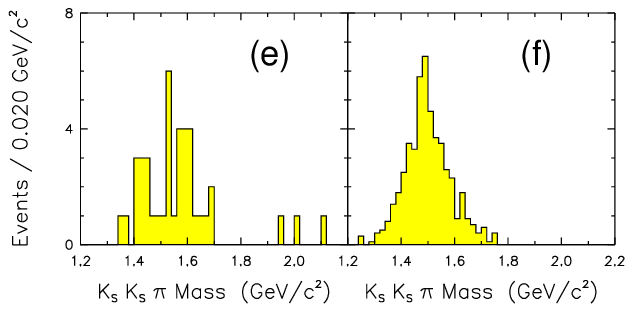}
  \includegraphics{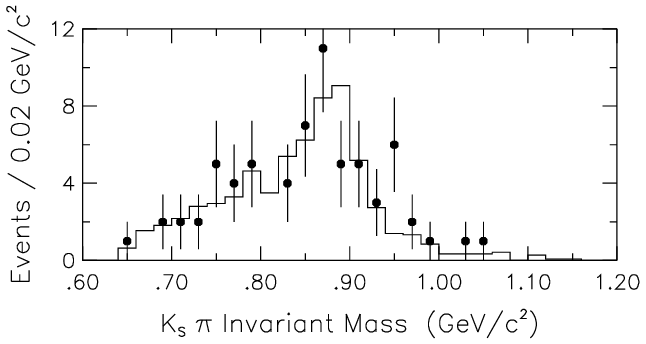}
  \end{picture}
  \caption[]{\small 
             (a) Plot
             of $M_{\pi^- K^0\pi^0}$ for  $\tau^-\!\!\rightarrow\!
             \nu_\tau h^- K^0_S\pi^0$ candidates in the data (points),
             along with a fit (solid histogram) to distributions obtained
             from $\tau^-\!\!\rightarrow\! \nu_\tau {K_1}^-(1270)$, 
             $\tau^-\!\!\rightarrow\! \nu_\tau {K_1}^-(1400)$, 
             and $\tau^-\!\!\rightarrow\! \nu_\tau K^- K^0 \pi^0$ 
             Monte Carlo samples (shown separately in (b), (c), (d)
             respectively), with the latter fixed according to the
             measured branching fraction.  In (e) and (f) the $K^0_S
             K^0_S \pi^-$ invariant mass is plotted for 
             $\tau^-\!\!\rightarrow\! \nu_\tau\pi^- K_S K_S$ candidate events
             in the data and Monte Carlo respectively, where the 
             latter is based on the model of Decker \emet \cite{decker}
             In (g), we plot the $K_S\pi^-$ submass (two entries per event)
             for data (points) and Monte Carlo.
            }
\label{fig:mkshpi0}
\end{figure}

Space constraints forbid a detailed discussion of the very interesting
3-meson final states containing one or two kaons, and only the results
can be presented below.  CLEO-II has observed the decays 
$\tau^-\!\!\rightarrow\!\nu_\tau h^- K^0_S \pi^0$, 
$\tau^-\!\!\rightarrow\!\nu_\tau K^- K^0_S \pi^0$, 
and $\tau^-\!\!\rightarrow\!\nu_\tau \pi^- K^0_S K^0_S$, the first two being
studied in a fashion similar to what was described in the previous 
section.  We obtain the following preliminary results:
\begin{eqnarray}
   B(\tau^-\!\!\rightarrow\!\nu_\tau h^-K^0\pi^0) 
                                      & = & 0.519\pm 0.035\pm 0.062\,\% \\
   B(\tau^-\!\!\rightarrow\!\nu_\tau K^-K^0\pi^0) 
                                      & = & 0.129\pm 0.050\pm 0.032\,\% \\
   B(\tau^-\!\!\rightarrow\!\nu_\tau \pi^-K^0\bar{K}^0) 
                                      & = & 0.083\pm 0.017\pm 0.017\,\%,
\end{eqnarray}
where the last result assumes that the $\pi^-K^0\bar{K}^0$ system 
manifests itself as $\pi^-K_S K_S$ 25\% of the time.
  The decay $\tau^-\!\!\rightarrow\!\nu_\tau \pi^- \bar{K}^0 \pi^0$ is thought
to proceed via the strange axial vector $K_1(1270)$ and $K_1(1400)$
resonances.  As shown in Figure \ref{fig:mkshpi0}(a-d), 
we find that both contribute, with a preliminary fit favoring
$K_1(1270)$ production, contrary to indications from TPC/$2\gamma$
\cite{tpc} in the all-charged channel, (although the admixture of 
$K^-K_S\pi^0$ events in our $h^-K_S\pi^0$ sample makes this difficult to
quantify).

  The resonant structure for the non-strange $[KK\pi]^-$ system is unknown.
Our results support crudely a recent model based on chiral perturbation
theory \cite{decker}, as shown in Figure \ref{fig:mkshpi0}(e-g) for the
$\pi^-K_S K_S$ event sample.
However the branching fraction for the $K^-K^0\pi^0$
channel and indications for substructure containing ${K^*}^0$ and ${K^*}^-$
in this channel are at variance with the predictions of this model.

%%%%%%%%%%%%%%%%%%%%%%%%%%%%%%%%%%%%%%%%%%%%%%%%%%%%%%%%%%%%%%%%%%%%%%%%%%%
%%%%%%%%%%%%%%%%%%%%%%%%%%%%%%%%%%%%%%%%%%%%%%%%%%%%%%%%%%%%%%%%%%%%%%%%%%%

\section{Conclusion}

To conclude, CLEO-II has obtained new results for many semi-hadronic
$\tau$ decay channels.  Many aspects of these processes are for the first
time being tested through precision measurements of branching fractions and
studies of spectral functions and resonant substructure.  We anticipate 
many more results from CLEO in years to come.
 
%%%%%%%%%%%%%%%%%%%%%%%%%%%%%%%%%%%%%%%%%%%%%%%%%%%%%%%%%%%%%%%%%%%%%%%%%%%
%%%%%%%%%%%%%%%%%%%%%%%%%%%%%%%%%%%%%%%%%%%%%%%%%%%%%%%%%%%%%%%%%%%%%%%%%%%

\section{Acknowledgements}

The CLEO collaboration acknowledges the effort of the CESR staff in 
providing excellent luminosity and running conditions.  This work was
supported by the U.S.~Dept.~of Energy and the National Science Foundation.

%%%%%%%%%%%%%%%%%%%%%%%%%%%%%%%%%%%%%%%%%%%%%%%%%%%%%%%%%%%%%%%%%%%%%%%%%%%
%%%%%%%%%%%%%%%%%%%%%%%%%%%%%%%%%%%%%%%%%%%%%%%%%%%%%%%%%%%%%%%%%%%%%%%%%%%

\bibliographystyle{unsrt}

%%%%%%%%%%%%%%%%%%%%%%%%%%%%%%%%%%%%%%%%%%%%%%%%%%%%%%%%%%%%%%%%%%%%%%%%%%%
\end{document}